\documentclass[sigplan,screen,nonacm,9pt]{acmart}
\usepackage{graphicx}
\usepackage{booktabs}
\usepackage{textcomp}
\usepackage{xcolor}
\usepackage{tikz}
\usepackage{enumitem}
\usepackage{multirow}
\usepackage{amsmath}
\usepackage{graphicx}
\usepackage{subcaption}
\usepackage{makecell}
\usepackage{amsmath}
\usepackage{mathtools}
\usepackage{amsthm}
\usepackage{hyperref}
\usepackage{soul}
\usepackage{xcolor}
\usepackage{colortbl}
\usepackage{booktabs}
\usepackage{algpseudocode}
\usepackage{algorithm}
\usepackage{setspace}
\usepackage{dblfloatfix}
\usepackage{arydshln}
\usepackage{gensymb}

\AtBeginDocument{%
  \providecommand\BibTeX{{%
    \normalfont B\kern-0.5em{\scshape i\kern-0.25em b}\kern-0.8em\TeX}}}

\begin{document}

\title{ Exploring and Exploiting Runtime Reconfigurable Floating Point Precision in Scientific Computing: a Case Study for Solving PDEs}

\author{Cong Hao, Georgia Institute of Technology, callie.hao@ece.gatech.edu}

\begin{abstract}

Scientific computing applications, such as computational fluid dynamics and climate modeling, typically rely on 64-bit double-precision floating-point operations, which are extremely costly in terms of computation, memory, and energy. While the machine learning community has successfully utilized low-precision computations, scientific computing remains cautious due to concerns about numerical stability. To tackle this long-standing challenge, we propose a novel approach to dynamically adjust the floating-point data precision at runtime, maintaining computational fidelity using lower bit widths.
We first conduct a thorough analysis of data range distributions during scientific simulations to identify opportunities and challenges for dynamic precision adjustment. We then propose a runtime reconfigurable, flexible floating-point multiplier (R2F2), which automatically and dynamically adjusts multiplication precision based on the current operands, ensuring accurate results with lower bit widths. Our evaluation shows that 16-bit R2F2 significantly reduces error rates by 70.2\% compared to standard half-precision, with resource overhead ranging from a 5\% reduction to a 7\% increase and no latency overhead. In two representative scientific computing applications, R2F2, using 16 or fewer bits, can achieve the same simulation results as 32-bit precision, while standard half precision will fail. This study pioneers runtime reconfigurable arithmetic, demonstrating great potential to enhance scientific computing efficiency.
Code available at \url{https://github.com/sharc-lab/R2F2}.
 
\end{abstract}

\maketitle

\section{Introduction}

\textbf{Scientific computing}, such as molecular dynamics, computational fluid dynamics, and climate modeling, aims at developing models and simulations to solve complex scientific and engineering problems. These applications usually require high numerical precision and predominantly employ double precision (64-bit) floating point operations. One reason is that in scientific computing, the data need to represent underlying physical phenomena as closely as possible and can thus be very large or small, requiring larger bitwidth to represent accurately. Such high-fidelity scientific computing is extremely computationally expensive and energy-consuming.

\begin{figure}
\centering
\begin{subfigure}{0.22\textwidth}
    \includegraphics[width=\textwidth]{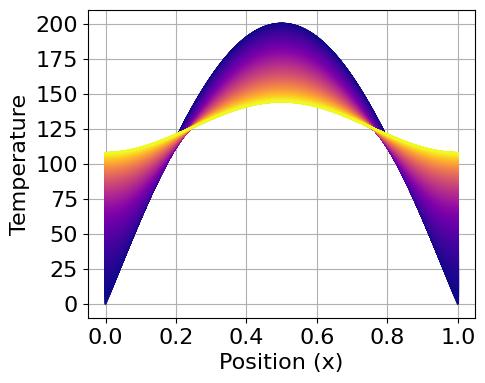}
    \caption{Use 32-bit floating point}
    \label{fig:plot-full-sin}
\end{subfigure}
\begin{subfigure}{0.22\textwidth}
    \includegraphics[width=\textwidth]{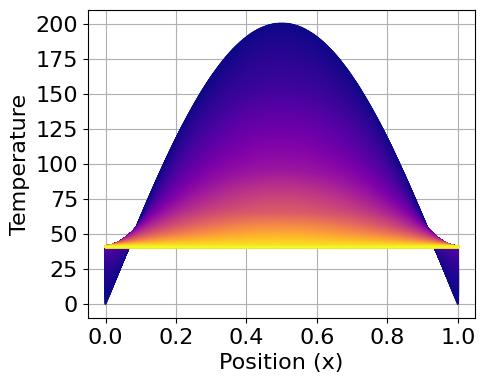}
    \caption{Use 16-bit floating point}
    \label{fig:plot-half-sin}
\end{subfigure}
\begin{subfigure}{0.22\textwidth}
    \includegraphics[width=\textwidth]{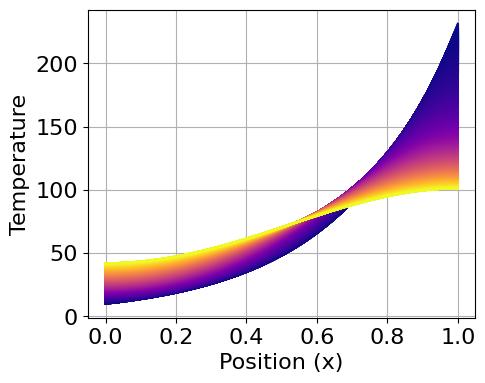}
    \caption{Use 32-bit floating point}
    \label{fig:plot-full-exp}
\end{subfigure}
\begin{subfigure}{0.22\textwidth}
    \includegraphics[width=\textwidth]{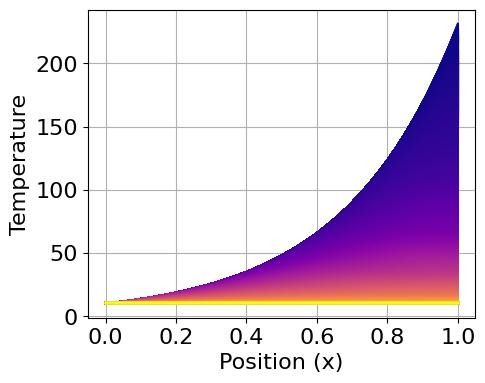}
    \caption{Use 16-bit floating point}
    \label{fig:plot-half-exp}
\end{subfigure}    
\caption{1D heat equation simulation using different data precision and heat initialization. (a)-(b): \texttt{sin}; (c)-(d): \texttt{exp} initialization. }
\label{fig:simulation-plot-32-vs-16}
\end{figure}

\textbf{Low precision computing} is a promising optimization to save computational energy, reduce memory traffic, enable larger cache blocks, and eventually reduce computing nodes in the supercomputer center. For instance, reducing precision from double to single (32-bit) can potentially reduce the number of computer nodes from 1000 to 500 with a significant saving. Despite the great success of using low precision in machine learning, the scientific computing community is more conservative. One major concern is whether an algorithm can remain numerically stable and achieve the same results as using higher precision. Fig.~\ref{fig:simulation-plot-32-vs-16} shows heat equation simulation results using single and half precision (16-bit in total and 5-bit for exponent; we denote it as E5M10), where the latter apparently results in wrong simulations.
Therefore, it remains a \textbf{great challenge} in scientific computing: \textit{how to reduce the computation precision while preserving high fidelity?}

Rich \textbf{pioneering studies} have been conducted to address this challenge. One popular approach is to use mixed precision in scientific computing~\cite{milroy2019investigating,lai2021mixed,paxton2022climate,abdulah2021accelerating,maynard2019mixed,chen2024mixed,strzodka2006pipelined}. Multiple studies have demonstrated that it is possible to use single precision partially, e.g., in a few subroutines, but not entirely, while using half precision will mostly fail~\cite{chen2024mixed,milroy2019investigating,lai2021mixed,paxton2022climate}.
These investigations are \textit{coarse-grained} at subroutine granularity and limited to standard precision types, while opportunities for using \textit{fine-grained and customized precision} remain unexplored, mainly due to the lack of hardware support.
On the other hand, approximate computing (AxC), e.g., AxC floating-point units~\cite{imani2017cfpu,imani2018rmac,imani2019approxlp,chen2021pam}, has demonstrated success in reducing energy consumption for machine learning applications. However, for scientific computing, since the possible data range is much wider than machine learning weights and activations, AxC units will still need a large bitwidth to accommodate this range, thus limiting the potential of using fewer bits. Additionally, non-deterministic approximation errors are prohibitive for high-fidelity scientific computing.

In this work, we tackle this challenge from a different angle than the traditional mixed-precision approaches: to distinguish, we name our scheme \textbf{runtime reconfigurable precision}, dubbed \textbf{rr-precision}. We ask the following research questions: (Q1) \textit{Are there opportunities to use lower bitwidth with rr-precision, i.e., is rr-precision indeed promising?} (Q2) \textit{If such opportunities exist, how can we exploit the benefit on hardware?}
To answer Q1, we explore the data range distribution during scientific simulations and discover that the data exhibit dynamic and local clusters, which are desirable features to exploit. To answer Q2, we propose a runtime reconfigurable floating point multiplier, which supports customized precision to maintain low bitwidth and accurate simulation results.
Our main contributions are summarized as follows:

\begin{itemize}[left=0pt]
    \item \textbf{Exploration -- Q1.} We are the first to conduct a fine-grained exploration for data range distribution in scientific computing using a case study, by developing an open-source library for floating point multiplications using arbitrary data precision. We successfully observe opportunities that have been overlooked: the data exhibits local clusters with dynamic range shift, advocating for mixed precision and precision adjustment at runtime.

    \item \textbf{Exploitation -- Q2.} We propose \textbf{R2F2}, a \textbf{\ul{R}}untime \textbf{\ul{R}}econ\textbf{\ul{F}}igurable \textbf{\ul{F}}loating point multiplier, whose precision can be automatically adjusted depending on the current operands.  
    R2F2 represents a floating point number using three sets of bits: fixed exponent bits, fixed mantissa bits, and flexible bits that can be allocated to either exponent or mantissa, decided at run-time by a lightweight precision adjustment unit.
    It does not rely on dynamic hardware reconfigurability (e.g., partial dynamic reconfiguration of FPGAs) and thus has high practicality.

    \item \textbf{Evaluation.} We implement R2F2 on FPGA and conduct thorough evaluation by comparing R2F2 with standard single and half multiplications.
    Comparing with half and smaller bitwidth, R2F2 achieves on average 70.2\%, 70.6\%, and 70.7\% error reduction using 16, 15, and 14 bits, respectively.
    Used in two real-world scientific computing applications, when half precision produces wrong simulation results, R2F2 using the same or even fewer bits is able to achieve the same simulation accuracy with no latency overhead. Compared to half precision, R2F2 utilizes 3\% to 7\% more LUTs (look-up tables), while the FF (flip-flop) overhead varies between a 5\% reduction and a 2\% increase. Compared to single precision, R2F2 reduces LUTs by 37.9\% and FFs by 33.2\%.

\end{itemize}

\section{Background and Related Work}

\textbf{Scientific computing.}
Scientific computing involves the development and application of computational algorithms and numerical methods to simulate physical phenomena, analyze data, and predict outcomes~\cite{golub2014scientific,heath2018scientific}, with an essential request of solving (partial) differential equations (PDEs)~\cite{golub1992scientific}. 
In this work, we use two representative PDEs as our case studies: the 1D heat equation solved using explicit finite difference scheme~\cite{zbMATH03905045}, and the 2D shallow water equations (SWEs) solved using Lax-Wendroff method~\cite{kinnmark2012shallow}. The former models heat conduction processes and the latter describes the behavior of shallow water bodies, accounting for the propagation of waves and fluid flow dynamics on earth.
The heat equation is expressed as $
\frac{\partial u(x,t)}{\partial t} = \alpha \frac{\partial^2 u(x,t)}{\partial x^2}$.
The SWEs are expressed as $\frac{\partial h}{\partial t} + \frac{\partial (hu)}{\partial x} + \frac{\partial (hv)}{\partial y} = 0$, $\frac{\partial (hu)}{\partial t} + \frac{\partial (hu^2 + \frac{1}{2}gh^2)}{\partial x} + \frac{\partial (huv)}{\partial y} = 0$, and $\frac{\partial (hv)}{\partial t} + \frac{\partial (huv)}{\partial x} + \frac{\partial (hv^2 + \frac{1}{2}gh^2)}{\partial y} = 0$.
We omit the variable definitions and numerical solutions due to space limit.

\noindent
\textbf{Mixed-precision in scientific computing.}
There have been long-lasting efforts in scientific computing using mixed precision operations, attempting to reduce, at least partially, double to single or half precision~\cite{milroy2019investigating,lai2021mixed,paxton2022climate,abdulah2021accelerating,maynard2019mixed,chen2024mixed,strzodka2006pipelined}. For example, Milroy et al.~\cite{milroy2019investigating} empirically explored the possibility of using mixed precision in the well-known Morrison-Gettelman microphysics package in climate modeling. They discovered that replacing all double with single leads to disastrous simulation results, while some subroutines are insensitive and can thus use single precision.
Chen et al.~\cite{chen2024mixed} conducted a similar exploration in the dynamical core of the Global-Regional Integrated Forecast System~\cite{li2023intercomparison}. Paxton et al.~\cite{paxton2022climate} studied the effects of deterministic and stochastic rounding in single and half precision and found that with stochastic rounding, 16-bit half precision may be useful in future climate modeling.

In contrast to such coarse-grained exploration, we propose \textit{fine-grained exploration that uses arbitrary precision} at the single-operation granularity. For example, we notice that reducing from 32-bit to 16-bit may be too aggressive since a 5-bit exponent is not sufficient, but using 15 bits in total with a 6-bit exponent may suffice.

\noindent
\textbf{Efficient hardware for floating point computing.}
Hardware community has also been striving to develop low precision computing units.
One line of effort is approximate computing (AxC) units.
For example, CFPU~\cite{imani2017cfpu}, RMAC~\cite{imani2018rmac}, ApproxLP~\cite{imani2019approxlp}, and PAM~\cite{chen2021pam}, Yin et. al~\cite{yin2016design} are representative AxC floating point multipliers;
Omidi et. al~\cite{omidi2021design}, Liu et. al ~\cite{liu2014inexact}, Liu et. al~\cite{liu2015design} are AxC adders.
Strzodka and Goddeke~\cite{strzodka2006pipelined} proposed mixed precision computing in iterative method using FPGAs.
Although promising energy reduction has been demonstrated, such approximation methods are highly empirical~\cite{imani2017cfpu,imani2018rmac,imani2019approxlp,chen2021pam}, or rely on the ad-hoc profiling of a specific algorithm~\cite{strzodka2006pipelined}, and thus lack generality.
In addition, due to the fixed precision format, they do not address the challenge of extremely wide data range with overflows.

In contrast to fixed precision units, we propose \textit{runtime reconfigurable} multiplication. 
Mu and Kim~\cite{mu2022dynamic} propose a promising PDE solver using dynamically reconfigurable precision, showing similar vision with ours; however, it only has bit-serial adders for integer addition using 4/8/12/16 bits, with no floating point operations and no discussion about how to adjust the precision at run-time.


\section{Exploration: Opportunities and Challenges of Dynamic Data Precision}

We first conduct an exploration to better understand scientific computational properties, aiming to reveal any overlooked opportunities and the challenges in exploiting them. We ask the following questions: 
\textbf{(1)} Are there special properties in the data that may provide opportunities to enable low-precision computation?
\textbf{(2)} If yes, what are the challenges to exploit them?

\subsection{Data Distribution Properties}
\label{sec:data-range}

\noindent
\textbf{Assumptions.}
Since scientific simulations mostly describe natural physical phenomena, an important assumption is that \textit{most values change smoothly and gradually}, such as heat conduction or diffusion and wave propagation through non-turbulent mediums. In our two case studies, we confirmed this scenario by visualizing intermediate and final data (e.g., Fig.~\ref{fig:simulation-plot-32-vs-16} and Fig.~\ref{fig:shallow-water}).
However, there are situations with possible sudden value changes, such as shock waves and water phase transitions.
Our focus in this study is primarily on simulations with smooth value changes, while our proposed approach can also handle sudden value changes, although less efficiently.

\noindent
\textbf{Data distribution properties.}
We analyze data distribution using the 1D heat equation during its entire simulation process.

\begin{figure}
\centering
\begin{subfigure}{0.49\textwidth}
    \includegraphics[width=\textwidth]{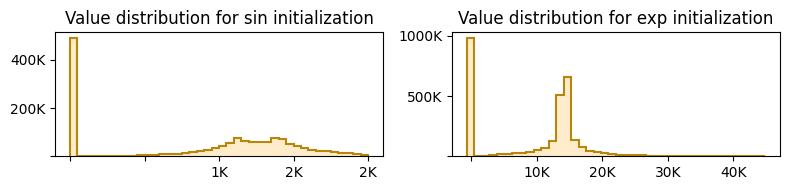}
    \caption{Distribution for all data values during the entire simulation process.}
    \label{fig:distribution-all}
\end{subfigure}
\begin{subfigure}{0.49\textwidth}
    \includegraphics[width=\textwidth]{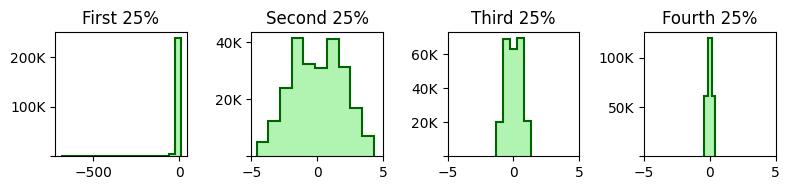}
    \caption{Distribution of small values at different stages of the simulation.}
    \label{fig:distribution-small-exp}
\end{subfigure}
\begin{subfigure}{0.49\textwidth}
    \includegraphics[width=\textwidth]{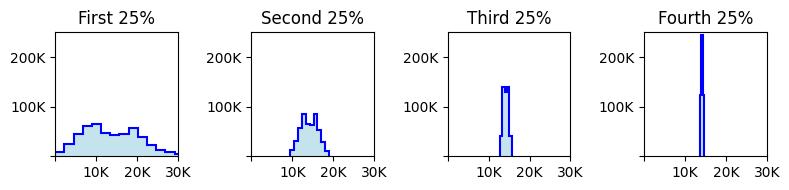}
    \caption{Distribution of large values at different stages of the simulation.}
    \label{fig:distribution-large-exp}
\end{subfigure}
\caption{Data distribution in the heat equation simulation. (a) implies that the data is globally wide while some clusters remain locally narrow; (b) and (c) demonstrate dynamic data range shift.}
\label{fig:data-distribution}
\end{figure}

\ul{First}, Fig.~\ref{fig:distribution-all} shows that the data values can have a wide range during the whole simulation: a massive cluster of small values with almost no change, and a wide range of large values. It reveals that the data range is \textbf{globally wide}, which is one reason to use double precision.
\ul{Second}, it also shows that data is locally clustered, and such clustering may also appear when simulation progresses.
Fig.~\ref{fig:distribution-small-exp} and Fig.~\ref{fig:distribution-large-exp} show that during the simulation, the data distribution changes and can exhibit strong clusters.
Fig.~\ref{fig:distribution-small-exp} is the distribution of small values; apparently, in the first 25\% simulation iterations, the smallest value can be -500; in the second and third 25\% iterations, the value range shrinks to (-5, 5) and (-1, 1); in the last 25\%, all values fall in (-0.25, 0.25). 
This observation reveals that the data range can be \textbf{locally narrow} and \textbf{dynamically changing}, and the clusters can emerge along with the simulation proceeds. This observation suggests that the \textit{data precision should be dynamically adjusted at runtime}, based on the current operands.
\ul{Third}, in the heat equation simulation, although using standard half precision (E5M10) leads to wrong results, using E6M9 for the multiplications whose operands are smaller than 0.0001 can compute correctly (as will also be shown in the experiment section). This observation strongly suggests \textit{customized data precision}.

\subsection{Optimal Data Precision: Are Intuitions Reliable?}
\label{sec:dynamic-precision-intuition}

Motivated by the observations in Sec.~\ref{sec:data-range}, it is preferable to dynamically determine the data precision. 
Intuitively, if two multiplication operands fall into value range $(v_{\text{min}}, v_{\text{max}})$ (assuming the two operands are within the same range), the optimal number of exponent bits should be:
\begin{equation}
n_{\text{expn}} =
\begin{cases}
    \lceil \log{(v^2_{\text{max}}) } \rceil + 1 & \text{if } v_{\text{max}} \geq 1 \\
    \lceil \log{((1/v_{\text{max}})^2) } \rceil + 1 & \text{if } v_{\text{max}} < 1
\end{cases}
\label{eq:emperical-exp-calc}
\end{equation}

\begin{figure*}
    \centering
    \includegraphics[width=0.94\textwidth]{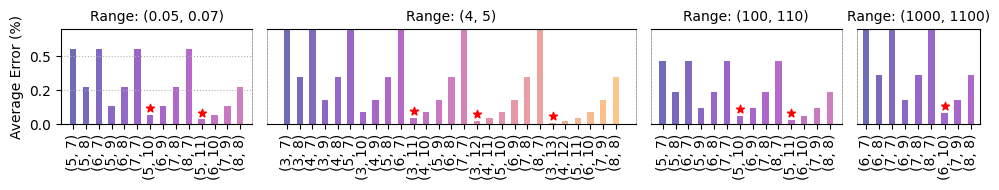}
    \caption{Average computation error using different configurations for floating point precision. X-axis shows the data precision in the format of (exponent, fraction); Y-axis is the average error percentage comparing with 32-bit floating point counterpart.}
    \label{fig:optimal-config-for-different-ranges}
\end{figure*}

However, despite the seemingly straightforward intuitions, in our experiments, we observe that dynamically determining the optimal data precision configuration in practice is non-trivial.
We conduct a comprehensive profiling for different precision configurations when multiplication operands fall into different ranges.
Fig.~\ref{fig:optimal-config-for-different-ranges} shows a subset of the profiling results. Apparently (and intuitively), different operand ranges prefer different precision.
For example, multiplications within range (0.05, 0.07) favor 5-bit exponent and 10/11-bit mantissa;
multiplications within range (4, 5) favor 3-bit exponent;
larger data ranges favor more exponent bits.

From this experiment, we notice that the empirical equation to compute the number of exponent bits does not always hold true. As shown in Fig.~\ref{fig:optimal-config-for-different-ranges}, Eq.~\eqref{eq:emperical-exp-calc} suggests 4 bits for exponent for operand in range (0.05, 0.07), while our profiling suggests 5 bits;
it suggests 6 and 8 bits for exponent in range (100, 110) and (1000, 1100), respectively, while our profiling suggests 5 and 6 bits.
Such discrepancy comes from various numbers of mantissa bits and the range difference between multiplication result and operands.
In addition, computing Eq.~\eqref{eq:emperical-exp-calc} can be expensive on hardware since it involves multiplication and logarithmic calculations, which contradicts the purpose of hardware efficiency.

Therefore, our exploration suggests that: (1) represent data using low bitwidth but flexible precision to accommodate diverse ranges; (2) adjust precision at runtime to accommodate range shift.


\section{Exploitation: Runtime Reconfigurable Multiplier with Precision Adjustment}


\subsection{Proposed Flexible Multiplier: R2F2}
\label{sec:proposed-mul}

We assume the two operands of a multiplication use the same precision; future work will consider different precisions.
Therefore, the multiplication is computed as XORing the
sign bits, adding the exponent bits, and multiplying the mantissa bits.

\begin{figure}
\centering
\begin{subfigure}{0.42\textwidth}
    \includegraphics[width=\textwidth]{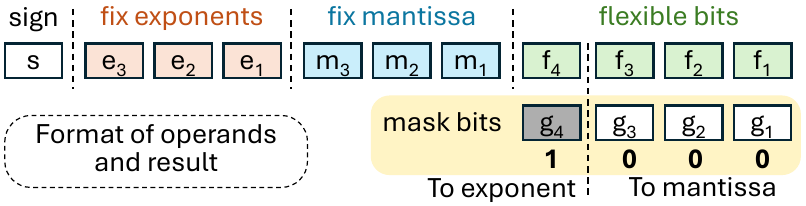}
    \caption{Flexible format}
    \label{fig:flex-format}
\end{subfigure}
\begin{subfigure}{0.47\textwidth}
    \includegraphics[width=\textwidth]{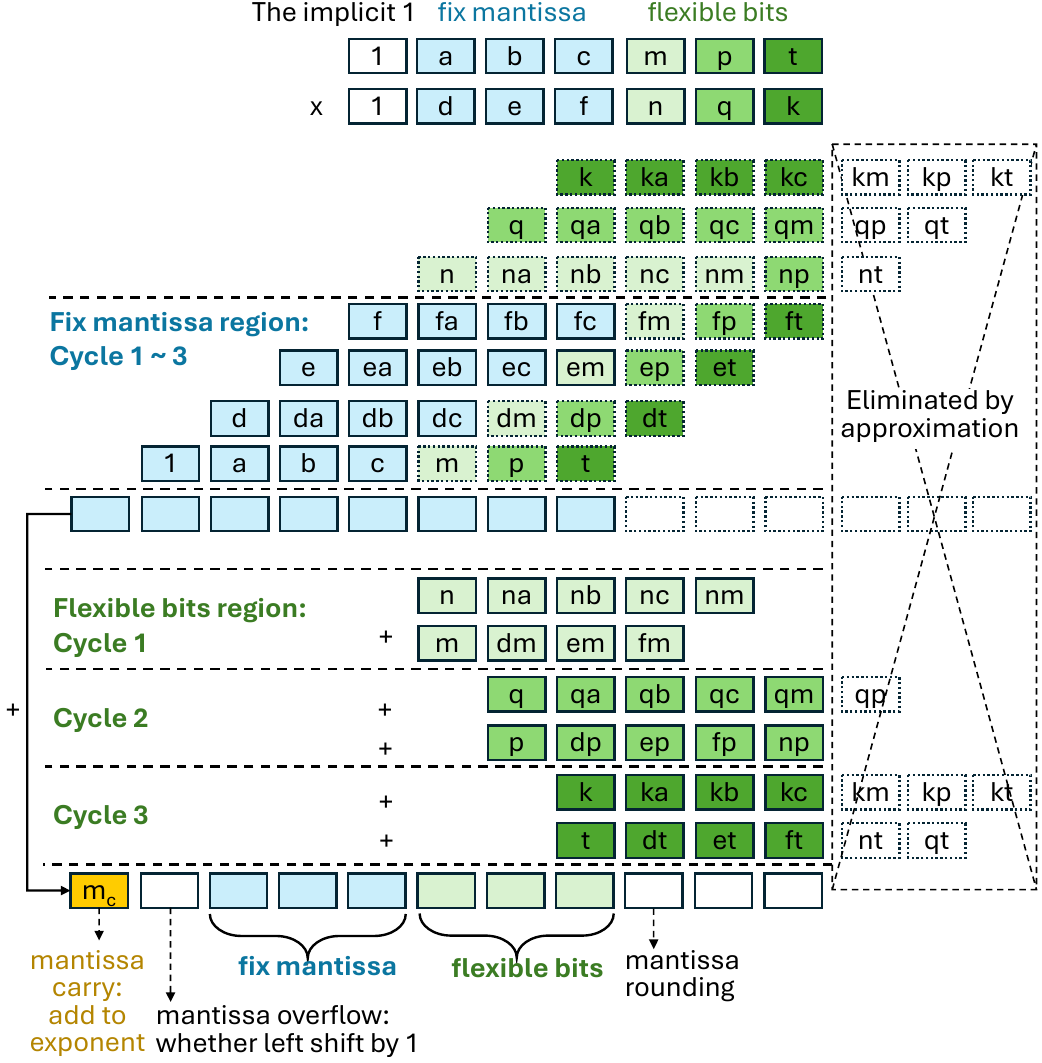}
    \caption{Mantissa bits computation}
    \label{fig:mant-compute}
\end{subfigure}
\begin{subfigure}{0.43\textwidth}
    \includegraphics[width=\textwidth]{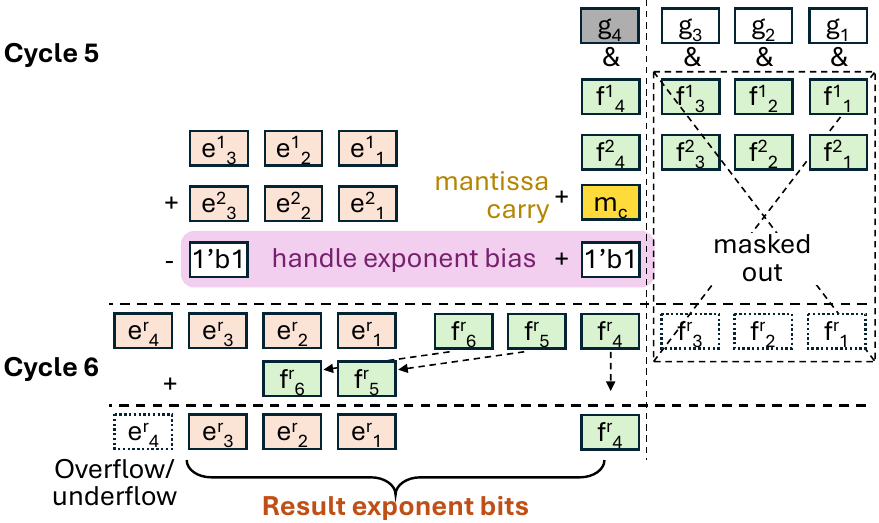}
    \caption{Exponent bits computation}
    \label{fig:expn-compute}
\end{subfigure}
\caption{Details of the proposed flexible floating point multiplier}
\label{fig:flexible-mul-design}
\end{figure}

\noindent
\textbf{Flexible data representation.}
The proposed flexible floating point data representation is shown in Fig.~\ref{fig:flex-format}.
We assume a fixed number of total bits, including four regions: one sign bit, a fixed exponent region with $EB$ bits, a fixed mantissa region with $MB$ bits, and a flexible region with $FX$ bits.
The flexible bits can be used to either represent exponent or mantissa, specified by $FX$ mask bits: a bit \texttt{1'b1} means that the corresponding flexible bit is used by exponent, while \texttt{1'b0} means mantissa. 
Omitting the sign bit, we denote the precision as <$EB$, $MB$, $FX$>.
Denoting $k$ as the bit count that will be allocated to exponent, $FX-k$ bits will be allocated to mantissa.

Apparently, in the proposed flexible representation, the exponent can use up to $EB+FX$ bits and thus has a larger dynamic range.
The standard half precision has 5 bits and can represent largest number 65504 ($2^{15} \cdot (1+1023/1024)$).
For R2F2, assuming the same 16 bits with precision <3, 8, 4>, the largest number can be represented is 1.8410715e+19 ($2^{63} \cdot (1 + 255/256) $), almost the same as 32-bit precision. When fewer bits are needed for exponent, more bits can be allocated for mantissa to improve numerical resolution.

Notably, the mask bits are not only used for distinguishing exponent and mantissa, but more importantly, reducing computing logic complexity and thus resource overhead. For instance, when adding exponents, as will be discussed later, directly adding two flexible regions using \texttt{AND} operations with the mask is more resource efficient than using large multiplexers to select and add the required bits only. It also avoids using expensive multiplexers for flexible bit selection operations, which could be abundant in F2R2.

\noindent
\textbf{Mantissa multiplication.}
Considering the implicit 1 before mantissa, 
the result mantissa is $1.m_1 \times 1.m_2$.
F2R2 computes the fixed mantissa and the flexible region simultaneously, where the flexible bits are computed one by one, one bit per cycle.
Fig.~\ref{fig:mant-compute} illustrates mantissa multiplication using an example with 3 fixed bits and 3 flexible bits for mantissa.
For ease of explanation, the first operand is denoted by $<1 ~ | ~ a ~ b ~ c ~ | ~ m ~ p ~ t >$ and the second operand is denoted by $<1 ~ | ~ d ~ e ~ f ~ | ~ n ~ q ~ k >$.
The fixed mantissa bits can be easily multiplied, i.e., $res \leftarrow 1abc \cdot 1def$, stored in $2\cdot MB$ bits.

Flexible bits are computed cycle by cycle and eventually added back to $res$, which requires additional storage bits in $res$. Rigorously, if $FX-k$ more bits are to be computed, then $2\cdot (FX-k)$ more bits are required, in addition to $res$. 
Since F2R2 allows arbitrary $ 0 \leq k \leq FX$, accommodating maximum possible value will introduce resource overhead of $2\cdot FX$ bits and require more computation, which defeats the purpose of low bitwidth.
Therefore, we approximate the results by only keeping $FX$ extra bits and eliminate the computation and storage after that, as shown in Fig.~\ref{fig:mant-compute}.
This approximation is valid since the rightmost $2\cdot FX$ bits will only contribute to mantissa rounding and eventually be discarded.
In our experiments, we validate that this approximate only introduces errors smaller than 0.1\% in less than 0.04\% of the time.

In the first cycle, flexible bits $m$ and $n$ are computed and stored in $res'$, where $res' \leftarrow (n \land op_1 + m \land op_2 ) \ll (FX-1) + (m \land n) \ll (FX-2) $. 
In the second cycle and afterwards, similar operations are executed but without the bits that are eliminated by approximation. For instance, in cycle 2, $res' \leftarrow res' + ( q \land op_1 + p \land op_2 ) \ll (FX-2) $; in cycle 3, $res' \leftarrow res' + k \land op_1 + t \land op_2$.
Finally, $res \leftarrow res \ll FX + res'$.
After all the bits are computed, mantissa is rounded and left shift if needed (i.e., if the highest bit is zero), and mantissa carry bit flagged accordingly, which will be added to exponent.

\noindent
\textbf{Exponent addition.}
Since exponent addition requires mantissa carry bit, we let exponent be computed after mantissa; in this example, it starts at cycle 5. However, this delay can be compensated if F2R2 is used in a pipelined manner.

Exponent addition is illustrated in Fig.~\ref{fig:expn-compute}.
In its first cycle, it computes fix mantissa bits and flexible bits simultaneously, including mantissa carry bit and excluding mantissa bits using the mask.
In its second cycle, it adds the two results together and sets the overflow/underflow bit.
Notably, since the exponent includes a bias, where $BIAS=2^{|e|-1}-1$ and $|e|=EB+k$, the result exponent is $e_{res} \leftarrow e_1 + e_2 - BIAS$.
Directly subtracting $BIAS$ from $e_{res}$ is not preferable since the value of $BIAS$ depends on $EB$ and the mask, and it requires complex logic to align the subtraction operation to the correct bits in the flexible region, given that $BIAS$ binary representation is all \texttt{1'b1}.
Alternatively, we notice that $e_{res}-BIAS = e_{res} -2^{|e|-1} + 1$, where the representation $2^{|e|-1}$ has only one leading \texttt{1'b1}, followed by all zeros, making the subtraction much simpler.
Better, the leading \texttt{1'1b} is always subtracted from the same bit in fix exponent region, which also simplifies the computation logic. In this example, $EB=3$ and $k=1$, so that $BIAS=2^{4-1}-1=7$, represented as \texttt{111 = 1000-1}. The leading bit \texttt{1'b1}, regardless of $EB$ and $k$ values, always aligns with the third bit in the fixed exponent region.

After mantissa and exponent computation, the individual intermediate results are assembled correctly to the final result registers.

\subsection{Dynamic Data Precision Adjustment}

\begin{figure}[t]
    \centering
    \includegraphics[width=0.42\textwidth]{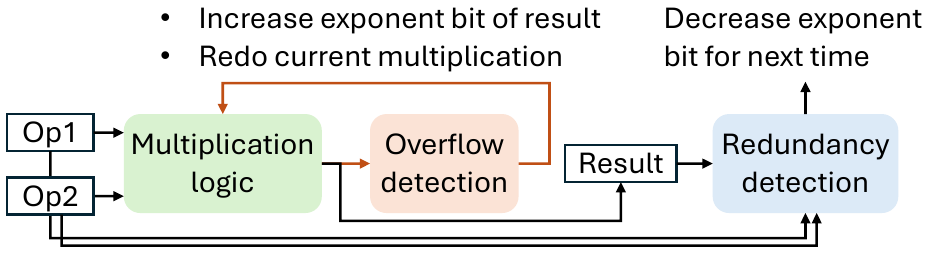}
    \caption{Logic to automatically adjust the precision.}
    \label{fig:auto_adjust}
\end{figure}

Motivated by the observations in Sec.~\ref{sec:dynamic-precision-intuition} that it is not straightforward and non-trivial to decide an optimal data precision, we propose a hardware-enabled lightweight precision adjustment unit, based on the data range of multiplication operands and the result.

The precision adjustment circuit has two functionalities: (1) increasing the exponent by one bit if an overflow or underflow is detected during computation, and (2) decreasing the exponent by one bit if a redundancy is detected in the operands and result. Mantissa bits will be decreased or increased accordingly, specified by the mask bits.
Fig.~\ref{fig:auto_adjust} illustrates the adjustment logic. If overflow/underflow is detected, together with increasing the exponent by one bit, it issues a signal to retry the multiplication using updated precision to store the result. 

Detecting overflow or underflow during the multiplication is trivial, as implemented in existing floating point multipliers~\cite{wang2010vfloat}.
To detect redundancy in exponent field, we examine the higher bits of the exponent. If the most significant bit (MSB) of exponent is zero, it represents values smaller than 1 since $e-BIAS$ will be negative, and MSB being one represents values larger than 1.
Following the leading MSB, consecutive zeros or ones indicate redundancy. For example, an 8-bit exponent \texttt{10000111} representing $2^{135-127}$ has the same value of a 5-bit exponent \texttt{10111} representing $2^{23-15}$.
The more zeros and ones, the larger redundancy.
Therefore, we examine the two bits after the leading MSB for redundancy identification; using one bit is too sensitive and may easily result in overflow in the reduced precision representation, and using three bits is too conservative which only works with more than four bits exponents.

\section{Experimental Results}


\noindent
\textbf{Experiment setup.}
The proposed R2F2 is implemented using High Level Synthesis (HLS), synthesized using AMD/Xilinx Vitis HLS 2023, and mapped to Pynq-Z2 FPGA board.
The two case studies, heat equation and shallow water equations, are implemented using C++ and computed with R2F2 using HLS. To distinguish, we denote R2F2 as $<EB, MB, FX>$ (see Sec.~\ref{sec:proposed-mul}), and the fixed floating point types as E5M10 (standard half with 5-bit exponent and 10-bit mantissa), or E5M9 and E5M8, similarly.

\subsection{Multiplication Accuracy}

\begin{figure}[t]
    \centering
    \includegraphics[width=0.48\textwidth]{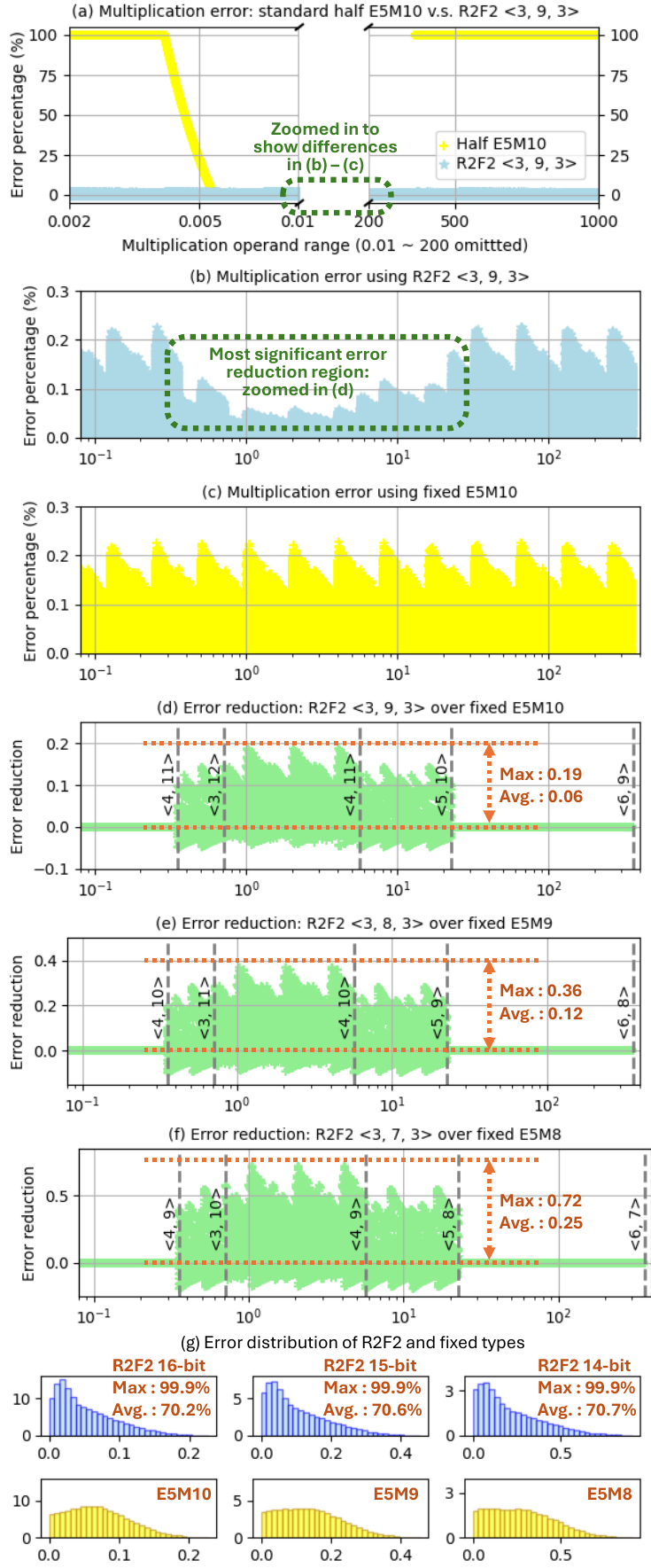}
    \caption{Error reduction of R2F2 compared with fixed types.}
    \label{fig:error-difference}
\end{figure}

We first evaluate 16-bit R2F2 with two baselines: standard single and standard half floating point multiplications, to demonstrate its superior accuracy.
We also evaluate different configurations of R2F2 and compare with its fixed precision counterpart to demonstrate the benefit of dynamic precision adjustment.
During testing, we sweep the range of (0.0001, 10000) for operands, divided into 10K intervals, and each interval has 1000 randomly sampled data pairs.
Fig.~\ref{fig:error-difference} illustrates the comparisons.
To clarify, we examine two types of multiplication errors: \textit{absolute error}, which compares with single precision multiplications;
\textit{relative error}, which compares the absolute error of R2F2 against its fixed type counterpart.

First, Fig.~\ref{fig:error-difference}(a) compares the multiplication error of 16-bit R2F2 in <3, 9, 3> and standard half (E5M10); the yellow dots are errors of standard half, and the blue dots are R2F2. It is obvious that for small and large values beyond the range that E5M10 can represent, unacceptable errors will occur due to overflow (for illustration purpose, errors are cast to 100\% if overflow happens).
On the other hand, using R2F2, it can dynamically adjust its exponent bitwidth to avoid the overflow and re-do the multiplications.

Second, Fig.~\ref{fig:error-difference}(b)-(c) show the zoomed-in region within range (0.01, 200) for R2F2 and E5M10, respectively, where the y-axis is the absolute error. Fig.~\ref{fig:error-difference}(d) shows the error difference by subtracting (b) from (c): larger values mean that R2F2 is more accurate than E5M10.
The \textit{absolute error reduction} is between 0.06\% to 0.19\% (compared with 32-bit), benefiting from the dynamically adjusted bitwidth for exponent and mantissa. 
It shows that with data range shift, the number of exponent bits changes, e.g., from 5 to 4 to 3, depending on the current operands; therefore, more bits are allocated to mantissa and thus better accuracy is achieved.
Similar observations are shown in Fig.~\ref{fig:error-difference}(e)-(f). In (e), with 15 bits, fixed type uses E5M9 and R2F2 uses <3,8,3>, where R2F2 has up to 0.36\% smaller absolute error. In (f), with 14 bits, fixed type uses E5M8 and R2F2 uses <3,7,3>, where R2F2 has up to 0.72\% smaller absolute error.
With fewer total bitwidth, R2F2 shows a more obvious advantage comparing with fixed precision.
Note that there are some negative values showing that R2F2 sometimes has worse accuracy; it is because of the approximation we applied during mantissa multiplication. The largest error difference is 0.09\%.

Third, Fig.~\ref{fig:error-difference}(g) show the error distributions of R2F2 and its fixed type counterparts, annotated with the \textit{relative error reduction} of R2F2 comparing with the fixed types. R2F2 has an average of 70.2\% and maximum 99.9\% error reduction comparing with E5M10, and an average of 70.6\% and 70.7\% comparing with E5M9 and E5M8.


\subsection{Resource and Latency Overhead}

\begin{table}[]
    \centering
    \small
    \setlength{\tabcolsep}{4pt}

    \caption{ Resource and latency overhead of R2F2. \textbf{Cnt}: resource count. \textbf{OH}: resource overhead. \textbf{Lat.}: cycle counts. \textbf{II}: initial interval. }
    
    \begin{tabular}{c|c|c|c|c|c|c}
    \toprule
    & \multicolumn{2}{c|}{\textbf{Flip Flops}} & \multicolumn{2}{c|}{\textbf{Look-Up Tables}} & \multicolumn{2}{c}{\textbf{Cycles}}\\ \cline{2-7}
    & Cnt & OH & Cnt & OH & Lat. & II \\
    \hline

    Lib. 64-bit FP (HLS)   & 2180 & - & 3264 & - & 30 & 11 \\
    Lib. 32-bit FP (HLS)    & 492 & - & 1438 & - & 24 & 5 \\
    Lib. 16-bit FP (HLS)   & 318 & - & 740 & - & 26 & 5 \\
\hline
    Impl. 64-bit FP     & 2032 & 2.82\texttimes  & 15650 & 3.20\texttimes & 13 & 4 \\
    Impl. 32-bit FP    & 1025 & 1.42\texttimes & 8093 & 1.66\texttimes & 13 & 4 \\
    \textbf{Impl. 16-bit FP}    & 720 & \textbf{1.0} & 4888 & \textbf{1.0} & \textbf{12} & \textbf{4} \\
\hline

    R2F2 16-bit <3,9,3>     & 710 & \textbf{0.99\texttimes} & 5161 & \textbf{1.06\texttimes} & 12 & 4 \\
    R2F2 16-bit <3,8,4>     & 720 & \textbf{1.00\texttimes} & 5132 & \textbf{1.05\texttimes} & 12 & 4 \\
    R2F2 16-bit <3,7,5>     & 731 & \textbf{1.02\texttimes} & 5152 & \textbf{1.05\texttimes} & 12 & 4 \\
\hdashline
    R2F2 15-bit <3,8,3>     & 696 & \textbf{0.97\texttimes} & 5091 & \textbf{1.04\texttimes} & 12 & 4 \\
    R2F2 15-bit <3,7,4>     & 713 & \textbf{0.99\texttimes} & 5082 & \textbf{1.04\texttimes} & 12 & 4 \\
\hdashline
    R2F2 14-bit <3,7,3>     & 685 & \textbf{0.95\texttimes} & 5028 & \textbf{1.03\texttimes} & 12 & 4 \\
    R2F2 14-bit <3,6,4>     & 702 & \textbf{0.96\texttimes} & 5249 & \textbf{1.07\texttimes} & 12 & 4 \\

    \bottomrule
    \end{tabular}
    
    \label{tab:resource}
\end{table}

\begin{figure}
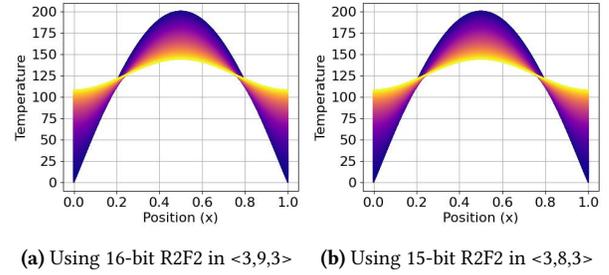

\centering
\begin{subfigure}{0.22\textwidth}
    \includegraphics[width=\textwidth]{figures/1D_heat_32_full_4K_sin.jpg}
    \caption{Using 16-bit R2F2 in <3,9,3>}
    \label{fig:heat-16bit}
\end{subfigure}
\begin{subfigure}{0.22\textwidth}
    \includegraphics[width=\textwidth]{figures/1D_heat_32_full_4K_sin.jpg}
    \caption{Using 15-bit R2F2 in <3,8,3>}
    \label{fig:heat-15bit}
\end{subfigure}
\caption{Heat equation simulation results using R2F2.}
\label{fig:heat-results-16-15}
\end{figure}

\begin{figure}
\centering
\begin{subfigure}{0.49\textwidth}
    \includegraphics[width=\textwidth]{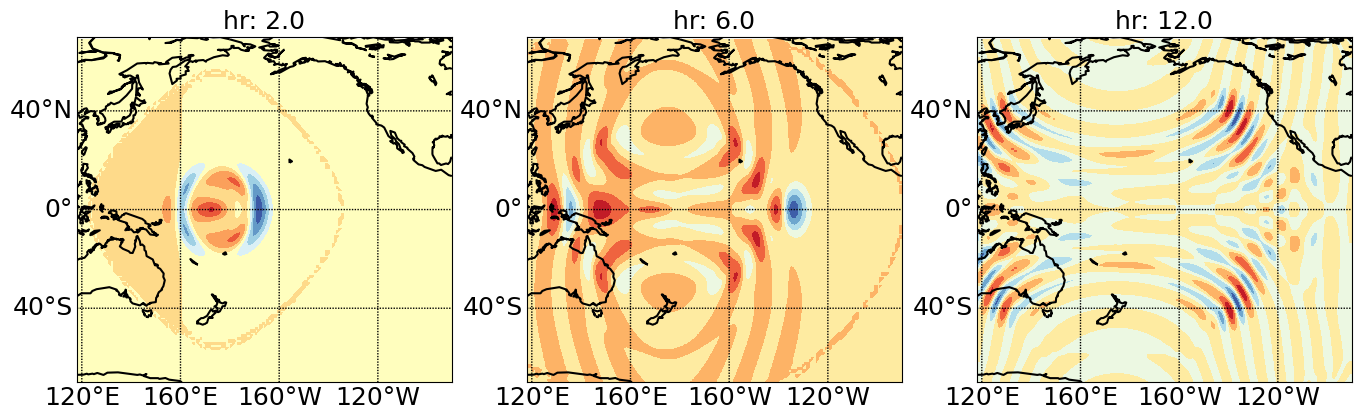}
    \caption{All sub-equations use 64-bit double precision}
    \label{fig:all-double}
\end{subfigure}
\begin{subfigure}{0.49\textwidth}
    \includegraphics[width=\textwidth]{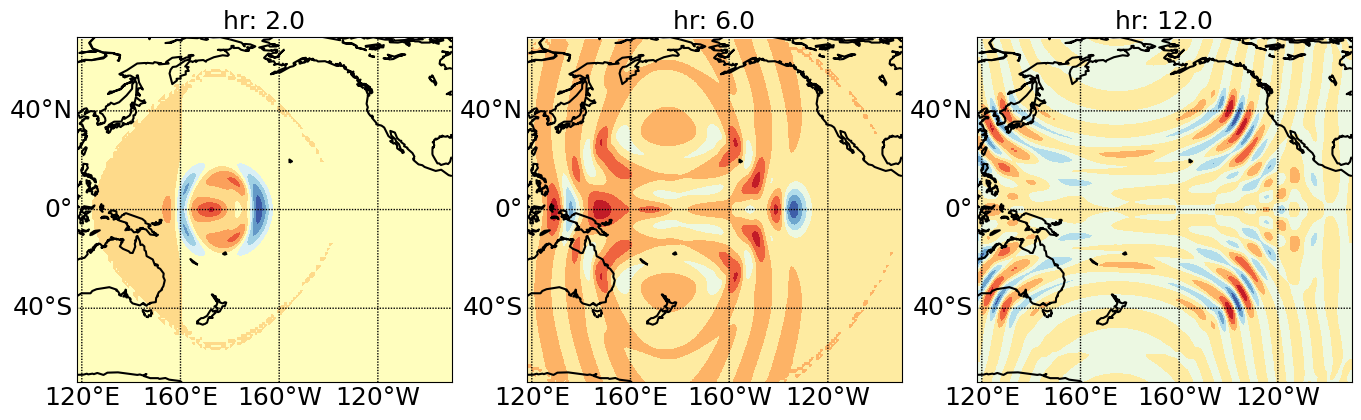}
    \caption{One sub-equation uses the proposed flexible 16-bit multiplier}
    \label{fig:all-flexible}
\end{subfigure}
\begin{subfigure}{0.49\textwidth}
    \includegraphics[width=\textwidth]{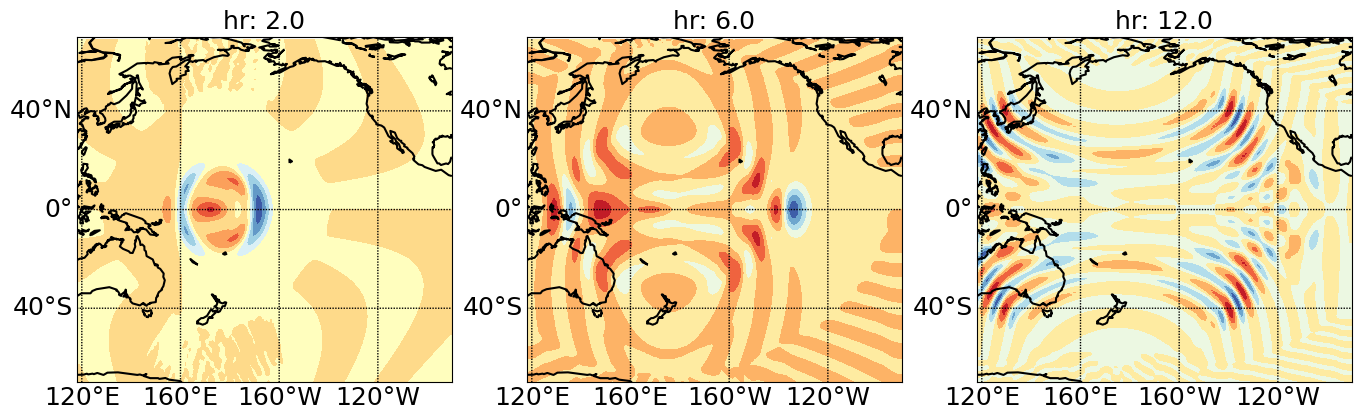}
    \caption{The same sub-equation uses standard fixed 16-bit multiplier}
    \label{fig:all-fixed}
\end{subfigure}
\caption{SWE simulation results using different precisions.}
\label{fig:shallow-water}
\end{figure}

With improved multiplication accuracy, the biggest question is how much resource and latency overhead R2F2 introduces.
Table~\ref{tab:resource} shows the comparison when implemented on FPGA. We deliberately enforce the tool not to use any DSPs so that Flip Flops (FFs) and Look-Up Tables (LUTs) can be a good proxy as circuit area.

We show two versions of double, single, and half multiplications: one uses Vitis HLS pre-designed library, which uses optimizations that are unknown to developers (row 1-3); the other is our own implementation in HLS (row 4-6), whose larger resource usage comes from peripheral logic such as type conversion. We emphasize that this provides a fair comparison baseline for the proposed R2F2, eliminating any hidden optimizations in Vitis HLS library.

Using the implemented E5M10 multiplier as the baseline, we evaluate R2F2 with different configurations.
For 16-bit R2F2 multipliers, there is 1\% reduction to 2\% increase in FFs and 5\% to 6\% in LUTs;
15-bit and 14-bit R2F2 multipliers show similar overhead ranging from 5\% to 1\% reduction in FFs and 3\% to 7\% increase in LUTs.
With almost negligible increase in LUT and even decrease in FF, 
R2F2 is able to achieve the same results as standard single precision, while standard E5M10 will fail in scientific simulation and thus is useless.

Table~\ref{tab:resource} also shows the latency overhead. It highlights that R2F2 can be pipelined in the same manner as standard multipliers do, and therefore has the same latency (12 cycles) and initial interval (II being 4 cycles).
The extra cycles besides those described in Sec.~\ref{sec:proposed-mul} include reading and converting from single precision to R2F2 format and converting back, where E5M10 does the same.

\subsection{Applied in Practice: Two Case Studies}

We apply R2F2 in the two scientific computing applications, heat equation and shallow water equations (SWEs).

Fig.~\ref{fig:heat-results-16-15} shows the heat equation simulation using (a) 16-bit R2F2 in <3,9,3> and (b) 15-bit R2F2 in <3,8,3> for \textit{all} multiplications, achieving the same simulation result as using single precision. 
Recall that Fig.~\ref{fig:plot-half-sin} shows that E5M10 will fail while R2F2 can succeed.
During the entire computation that involves 1.5M multiplications, R2F2 precision adjustment because of overflow happened only 5 times and thus introducing negligible re-run overhead;
precision adjustment because of redundancy happened 23 times.

Fig.~\ref{fig:shallow-water} shows the SWE computation results. The original SWE involves 24 equations and all use double precision. 
Since replacing all of them using single will fail, 
we only substitute the multiplications in one equation: 
$Ux_{mx}[i][j] = q1_{mx}[i][j] \cdot q1_{mx}[i][j] / q3_{mx}[i][j] + 0.5 g \cdot q3_{mx}[i][j] \cdot q3_{mx}[i][j]$.
Fig.~\ref{fig:shallow-water}(a) are the results using double precision after 2, 6, and 12 hours;
(b) are the results using 16-bit R2F2, and (c) are the results using E5M10.
Apparently, using E5M10 has inaccurate results: obvious in the first two plots and noticeable in the third (within \(120\degree\)E to \(40\degree\)S and \(160\degree\)E and \(160\degree\)W).
In contrast, using R2F2 leads to the same simulation results as double precision.
Within the 30K multiplications, R2F2 adjusted precision 7 and 15 times, because of overflow and redundancy, respectively.


\section{Conclusions}

This paper addresses the challenge of reducing data precision in scientific computing without compromising fidelity. By analyzing data range distributions, we identified opportunities for dynamic precision adjustment. Our proposed Run-time Reconfigurable Floating point multiplier (R2F2) effectively adapts precision at runtime based on data characteristics. R2F2 reduces error rates by 70.2\% on average compared to half precision, with negligible resource overhead. In practical applications, R2F2 achieves comparable simulation accuracy as double by using as few as 16 bits when half precision fails. R2F2 demonstrates its potential to revolutionize precision management in scientific computing with low bitwidth for energy and memory reduction.


\end{document}